\documentclass[twocol]{ametsocV6.1}
\usepackage{xcolor, algorithm, algorithmic}
\usepackage{booktabs}
\usepackage{bbm}

\title{Improving Ensemble CAPE Forecasts with a Diffusion Model Incorporating Aerosol Information}

\authors{Zachary James,\aff{a}\correspondingauthor{Zachary James, zj37@cornell.edu}
Joseph Guinness,\aff{b}
Arthur DeGaetano\aff{c}
}

\affiliation{\aff{a}{Cornell University, Department of Statistics and Data Science}\\
\aff{b}{Washington University in St. Louis, Department of Statistics and Data Science}\\
\aff{c}{Cornell University, Department of Earth and Atmospheric Sciences}}

\abstract{Convective available potential energy (CAPE) is an important variable for forecasting severe weather and understanding deep convection and precipitation. The latest versions of the Global Forecast System (GFS) and related Global Ensemble Forecast System (GEFS) have exhibited a bias towards underestimating CAPE values during the summertime. We train an artificial intelligence (AI) diffusion model to improve the skill and uncertainty quantification of afternoon 6-hour lead time ensemble forecasts over the United States. Our model takes a GFS CAPE forecast as input and outputs an ensemble that significantly outperforms both GFS and GEFS 6-hour forecasts on root mean square error, continuous ranked probability score, and Brier score. We propose a two-stage training pipeline to leverage both a larger historical GFS forecast dataset and a smaller historical GEFS dataset, despite the two using initialization and parameterization schemes that vary over time. We also show that classifier-free guidance can be used to control the skill and spread of the forecasts. We then demonstrate the versatility of our framework by adding aerosol optical depths (AODs) of black carbon, organic carbon, dust, sea salt, and sulfates as additional input features. Aerosols can invigorate or suppress convection depending on atmospheric conditions. Our AI models effectively incorporate aerosols to produce improved CAPE forecasts. We interpret the model components by using permutation feature importance to rank the influence of the different AODs and find that black carbon, organic carbon, and sulfate aerosols have a greater impact on the model's CAPE predictions than sea salt and dust aerosols.}

\begin{document}

\maketitle


\statement

Convective available potential energy (CAPE) is a key diagnostic variable for understanding weather conditions. High CAPE values may be indicative of strong convective systems and severe weather. However, the Global Forecast System and Global Ensemble Forecast System both struggle to forecast high CAPE values, particularly during the active summer season. We found that we can produce an ensemble forecast that outperforms GEFS by creating an artificial intelligence (AI) diffusion model that takes a GFS CAPE forecast as input. We then further improved the model by providing aerosol information, which is known to impact CAPE.

%

\section{Introduction}
\label{sec:intro}

Atmospheric convection is a fundamental part of the Earth system and plays a central role in forecasting and modeling phenomena such as precipitation and severe weather. Convective available potential energy (CAPE), which measures the amount of energy in the atmosphere available for convection, is one of several important variables for characterizing convective environments. CAPE is used in convective parameterization in numerical weather prediction (NWP) models \citep{Kain1993,Randall1997}, as a diagnostic for severe weather and thunderstorms \citep{Kunz2007, Kaltenbock2009}, for analyzing precipitation \citep{Lepore2015, Dong2019}, and for assessing changes in convective environments under climate change \citep{Ye1998}.

CAPE measures the buoyant energy available to a parcel of air if it rises adiabatically without mixing with the environment. It is calculated as 
\begin{equation}
    \text{CAPE} = \int_{z_{LFC}}^{z_{EL}} g\left(\frac{T_\mathrm{v}^{\text{parcel}}(z) - T_\mathrm{v}^{\text{env}}(z)}{T_\mathrm{v}^{\text{env}}(z)}\right)dz
\end{equation}
a vertical integral of parcel buoyancy from the height $z_{LFC}$ of the level of free convection (LFC) to the height $z_{EL}$ of the level of equilibrium (EL), where $T_\mathrm{v}^{\text{parcel}}$ and $T_\mathrm{v}^{\text{env}}$ are the virtual temperature of the parcel and environment respectively, and $g$ is Earth's gravitational acceleration . The LFC is the lowest level at which a parcel of air is positively buoyant relative to the environment, while the EL is the level above the LFC at which the parcel returns to zero buoyancy.

Forecasting summertime CAPE over CONUS is challenging. The lack of strong synoptic-scale forcing results in convection often being driven by meso-$\alpha$ scale features such as fronts and drylines \citep{Xia2024}. As a result, small initialization errors in thermodynamic parameters can lead to significant errors in CAPE. Most major NWP models, including NOAA's Global Forecast System (GFS), forecast CAPE as a diagnostic variable computed from the thermodynamic state. The last two versions of GFS, v15.1 and v16.0, have both been shown to have a bias towards underestimating CAPE values in the summertime over the coterminous United States (CONUS), when high CAPE environments are most prevalent. This bias is similarly seen in the Global Ensemble Forecast System (GEFS) \citep{Malloy2025}, a prediction system based on the dynamical core and associated parameterization schemes of GFS. 

The modeling and computational challenges of NWP have led to interest in artificial intelligence (AI) forecasting models. AI models use historical data to learn how variables evolve in space and time. Deterministic AI methods emulate NWP by mapping the current atmospheric state to a future state. For example, the GraphCast model from Google DeepMind uses a graph neural network to produce a global $0.25^\circ$ resolution forecast that outperforms the ECMWF's HRES forecast on multiple variables at different lead times \citep{Lam2023}. GraphCast is now offered on an experimental basis by ECMWF. 

Probabilistic AI methods, such as diffusion models (DM), try to produce a predictive distribution from which an ensemble can be sampled. DMs have become popular because of their ability to sample from complex distributions, particularly those of images \citep{Ho2020}. DMs learn the underlying distribution by training a model to recover a forecast that has had noise added to it. The final model is then used to iteratively denoise a sample of Gaussian noise to produce an actual forecast. DMs can sample from any distribution whose support is on a manifold. DeepMind's GenCast uses a DM to sample forecasts on a mesh. The resulting model produces global $0.25^\circ$ resolution ensemble forecasts that outperformed ECMWF's ENS forecast on multiple variables and lead times, and demonstrates skill in predicting tropical cyclone tracks \citep{Price2025}. Google Research's SEEDS-GPP uses a DM to emulate the full GEFS forecast given two randomly selected input forecasts \citep{Li2024}.  By using the ground truth as the response, they are able to produce an ensemble that outperforms GEFS.

Motivated by these results, we develop DMs to produce ensemble forecasts for CONUS during the extended summertime, defined as April 1 to September 30, that outperforms the GEFS CAPE forecast. We use a U-Net with attention as the denoiser and sample as a denoising diffusion probabilistic model (DDPM) \citep{Ho2020}. The models are trained to produce $0.5^\circ$-resolution 6-h ensemble forecasts issued at 00Z of summertime CAPE over CONUS, defined as the region bounded by $127.25^\circ$W–$63.25^\circ$W and $23.75^\circ$N–$55.75^\circ$N. We use the GEFS control member 0-h forecast at 00Z as a proxy analysis for evaluating model performance because it uses the same data assimilation scheme as the remaining GEFS forecast members and allows for a fair comparison between GEFS and the AI models. 

Since GEFS CAPE forecasts prior to October 2020 only have $1^\circ$ resolution, we do not have a sufficient number of historical forecasts to both train and test the model on GEFS data alone, so we propose a two-stage training procedure. First, we train our models on a base dataset that uses GFS 0-h forecasts at 00Z as the response. We then continue training on a fine-tuning dataset that uses the GEFS 0-h forecast at 00Z as the response. The final forecasts are calibrated in terms of skill and spread with classifier-free guidance. Our resulting models can produce a 30-member ensemble forecast for CONUS in approximately five minutes on an NVIDIA A40 GPU. We find that our models significantly outperform both GFS and GEFS in root mean square error (RMSE), continuous ranked probability score (CRPS), and Brier score.

We then demonstrate the versatility of our model by incorporating additional variables. Atmospheric aerosols are known to influence CAPE, and in turn, convection \citet{Jiang2018}. However, the relationship is complex, with multiple conflicting processes at play. When aerosols are present near the top of the planetary boundary layer (PBL), they may enhance stability within the PBL while increasing buoyancy above it, resulting in higher CAPE \citep{Wang2013}. Conversely, high AOD can result in lower CAPE through radiative effects such as decreased surface heating \citep{Jiang2018}. Aerosols can also affect CAPE indirectly through cloud microphysics. For example, higher concentrations of cloud condensation nuclei can increase cloud albedo, reducing surface heating \citep{Twomey1974}. We found that adding MERRA-2 average aerosol optical depth (AOD) data for five primary aerosols increased the accuracy of the forecasts, suggesting that the model was able to learn the relationship between the aerosols and CAPE. Furthermore, we were able to use permutation feature importance (PFI) to determine that black carbon, organic carbon, and sulfates have a greater impact on model prediction than dust and sea salt.

The remainder of this paper is presented as follows: Section \ref{sec:data} details NWP models and data sources used in this study. Section \ref{sec:dm} describes the DM, including its architecture, how it is trained, and how ensembles are produced. We also introduce our two-stage training pipeline and describe how we use classifier-free guidance to calibrate the forecasts. In Section \ref{sec:results} we present our results. Section \ref{sec:interpetation} interprets the model with PFI and ranks the aerosols by their impact on forecasts. We conclude in Section \ref{sec:conclusion} with a short discussion.

\section{Data and Numerical Models}
\label{sec:data}

\subsection{GFS}

GFS is a global NWP model that produces atmospheric forecasts at $0.5^\circ$ horizontal resolution, with lead times of up to 16 days. GFS has existed in various forms for several decades, with the latest version, GFS v16, becoming operational in March 2021, replacing GFS v15.1, which became operational in June 2019. Forecasters have long-noted that GFS tends to underestimate CAPE, which appears to have worsened in v15.1 and v16 \citep{forecasters2021}. This bias is particularly pronounced in GFS v16 summertime forecasts, which has been attributed to excessive mixing in the PBL, potentially the result of a dry soil moisture bias \citep{Xia2024}. 

\subsection{GEFS}

GEFS is a global ensemble forecasting system based on GFS. We are only concerned with the latest version, GEFS v12, which became operational in October 2020 \citep{emc2024gefs}. GEFS produces ensemble forecasts with 30 perturbed members and one control member. GEFS v12 uses the same dynamical core and physical parameterization schemes as GFS v15.1, and improves upon previous GEFS versions with $0.5^\circ$ resolutions and updated perturbation schemes. Similar to GFS, GEFS $0.5^\circ$ CAPE forecasts exhibit a bias towards low CAPE values, which is particularly noticeable in the summertime. Attempts to alleviate this bias, such as by scaling CAPE values, have found limited success \citep{Malloy2025}. 

\subsection{MERRA-2}

MERRA-2 is a global atmospheric reanalysis product that assimilates satellite and ground-based observations. We use MERRA-2 to obtain hourly-averaged AOD at 550 nm for black carbon, dust, organic carbon, sea salt, and sulfates \citep{merra2}. The AOD at wavelength $\lambda$ 
\begin{equation}
    \tau_A(\lambda) = \tau(\lambda)-[\tau_R(\lambda)+\tau_O(\lambda)+\tau_N(\lambda)]
\end{equation}
is calculated as the total optical depth $\tau(\lambda)$ minus the optical depth due to Rayleigh molecules $\tau_R(\lambda)$, ozone $\tau_O(\lambda)$, and nitrogen dioxide $\tau_N(\lambda)$. The AOD is then partitioned into the contributions of each of the five aerosols. MERRA-2 provides AOD at a $0.5^\circ \times 0.625^\circ$ resolution, so we use bilinear interpolation to downscale it to $0.5^\circ\times 0.5^\circ$.

\section{Probabilistic Diffusion Forecasts for CAPE}
\label{sec:dm}

\begin{sidewaystable*}[p]
\centering
\begin{tabular}{lllp{2.5cm}lp{3.5cm}}
\toprule
Model Name & Input Model & Period & Predictors & Response & Description \\
\midrule
CAPE-Only Base & --- & 2008--2020 & GFS 0-h Issued 18Z, GFS 6-h Issued 18Z& GFS 0-h Issued 00Z & Base model using GFS response, no aerosols, no fine-tuning\\
\addlinespace
CAPE-Only Finetuned & CAPE-Only Base & 2021--2022 & GFS 0-h Issued 18Z, GFS 6-h Issued 18Z  & GEFS 0-h Issued 00Z & CAPE-Only Base model finetuned on 2021--2022 GEFS responses\\
\addlinespace
Aerosol GFS Base & --- & 2008--2020 & GFS 0-h Issued 18Z, GFS 6-h Issued 18Z, MERRA-2 Aerosols& GFS 0-h Issued 00Z & Base model using 2008--2020, GFS response, with aerosol predictors \\
\addlinespace
Aerosol GEFS Base & --- & 2021--2022 & GFS 0-h Issued 18Z, GFS 6-h Issued 18Z, MERRA-2 Aerosols & GEFS 0-h Issued 00Z & Base model using 2021--2022, GEFS response, with aerosol predictors\\
\addlinespace
Aerosol Partially Finetuned & Aerosol GFS Base & 2021 & GFS 0-h Issued 18Z, GFS 6-h Issued 18Z, MERRA-2 Aerosols & GEFS 0-h Issued 00Z & Aerosol GFS Base finetuned using 2021 GEFS responses\\
\addlinespace
Aerosol Finetuned & Aerosol GFS Base & 2021--2022 & GFS 0-h Issued 18Z, GFS 6-h Issued 18Z, MERRA-2 Aerosols & GEFS 0-h Issued 00Z & Aerosol GFS Base finetuned using 2021--2022 GEFS responses\\
\bottomrule
\end{tabular}
\caption{Summary of statistical model configurations. Three of the models, CAPE-Only Finetuned, Aerosol Partially Finetuned, and Aerosol Finetuned, use an input model for the two stage training described in Section 3\ref{sec:twostage}. The other three are base models. The GFS and GEFS predictors and response consisted of only the CAPE forecasts.}
\label{tab:models}
\end{sidewaystable*}

\subsection{Models}

We aim to generate an ensemble forecast of CAPE over CONUS for 00Z on a target day, conditional on information available at 18Z: the GFS 0-h forecast issued at 18Z, the GFS 6-h forecast issued at 18Z, and optionally, MERRA-2 AOD. The response used during training is either the GFS 0-h forecast issued at 00Z or the GEFS 0-h control member forecast issued at 00Z, depending on the model. Some of our models will undergo a two-stage training procedure, outlined in Section 3\ref{sec:twostage}. All models are validated out of sample against the GEFS 0-h control member forecast issued at 00Z. We use the GEFS 0-h as a proxy for ground truth because it uses the same data assimilation scheme as the perturbed members, thus allowing for a fair comparison between GEFS and the AI models.

First, we train a model that does not have access to aerosol information. The model uses the GFS 6-h forecast issued at 18Z and the GFS 0-h forecast issued at 18Z to produce an ensemble 6-h forecast for the conditions at 00Z. The 0-h forecast serves a proxy for the CAPE conditions at 18Z, and we refer to it as the model initialization. Including this initialization helps the model learn the relationship between the initial CAPE conditions and subsequent forecast errors. Depending on the stage in the training pipeline, whose motivation is described more fully in Section 3\ref{sec:twostage}, we use either the GFS 0-h forecast issued at 00Z or the GEFS 0-h control member forecast issued at 00Z as the response during training. This produces two models: the CAPE-Only Base model, and the further trained CAPE-Only Finetuned model. The CAPE-Only models simultaneously perform two tasks: generating an ensemble from a deterministic forecast and enhancing the forecast, and are akin to the SEEDS-GPP model \citep{Li2024}. 

We then further improve our forecasts by incorporating aerosol information. In addition to the two inputs of the CAPE-Only models, the aerosol models use 17Z-18Z hourly-averaged AOD for the five primary aerosols from MERRA-2. This introduces information unavailable to GFS and GEFS, which do not have aerosol inputs. We chose to incorporate aerosols because they are known to have a complex relationship with thermodynamics and CAPE \citep{Jiang2018}. However, under our framework, any spatial variable can be added, even if the relationship with the forecast variable is unknown. The architecture of the aerosol models differ from the CAPE-Only models only in the first layer of the U-Net, which uses convolutional filters with five additional input channels. This results in a modest increase in the number of parameters. We again use a two stage training pipeline, resulting in more than one model. We train several different models with aerosols to test the efficacy of our two-stage training pipeline. These models are described in Section 3\ref{sec:twostage} and Table \ref{tab:models}.

\subsection{Diffusion}

Unlike a traditional deterministic NWP model, we treat the evolution of CAPE as a stochastic process. The future CAPE $Y$ given current atmospheric conditions and deterministic input forecast $X$ follows an unknown distribution $D\mid X$. $Y$ and $X$ represent conditions over the entire spatial domain and, at our resolution, are represented as vectors of dimension $m=61\times 119=7259$. Since the spatial domain is not global, a spherical representation is not necessary. We seek to draw $K$ samples $\{y_i\}_{i=1,\ldots,K}$ from $D\mid X$, which constitute the prediction ensemble. 

We train a conditional DDPM to sample from $D\mid X$ \citep{Ho2020}. While a multitude of more efficient DM solvers have been developed since DDPM \citep{Song2022, Lu2022, Lu2025}, many involve trade-offs between accuracy and efficiency, and we have chosen to prioritize predictive performance. Even then, the inference cost of our DM model is insignificant when compared to the cost of the underlying NWP model used as input.

For given attributes $x$, we assume that there exists a sequence of latent variables $\mathbf{Z}=\{Z_t\}_{t=1,\ldots,T}$, with $Z_0\sim D\mid X=x$ and $Z_T\sim N(0,I_m)$, that form a Markov chain with forward process
\begin{equation}
    Z_{t}\mid Z_{t-1} \sim N(\sqrt{1-\beta_t}Z_{t-1}, \beta_tI_m)
\end{equation}
and reverse process
\begin{equation}
    Z_{t-1}\mid Z_t \sim N(\mu_\theta(Z_t,x,t), \sigma_t^2I_m)
\end{equation}
where $\{\beta_t\}_{t=1,\ldots,T}$ and $\{\sigma^2_t\}_{t=1,\ldots,T}$ are predefined schedules detailed in Appendix A. The corresponding joint likelihood of the reverse process is intractable, so we estimate $\theta$ by optimizing an evidence lower bound (ELBO)
\begin{multline}
\label{eq:elbo}
    \log p_\theta(\mathbf{z}) \geq \mathbb{E}_q\Big[\log p_\theta(z_0\mid z_1) \\
    - \sum_{t=2}^T D_{KL}\!\left(q(z_{t-1}\mid z_t,z_0)\parallel p_\theta(z_{t-1}\mid z_t)\right)\Big]
\end{multline}
where $q$ and $p_\theta$ are the densities of the forward and reverse processes respectively, and $D_{KL}$ is the Kullback–Leibler divergence. \citet{Ho2020} optimized this objective stochastically by selecting a different timestep $t$ at each iteration of gradient descent. They also improved performance by parameterizing the reverse process in terms of the noise $\epsilon$ instead of the mean $\mu$. This yields the reweighted objective

\begin{multline}
    \mathcal{L}(\theta) = \mathbb{E}_{t\sim\text{Unif}\{1,\ldots,T\}}\Big[\lVert \epsilon \\
    - \epsilon_\theta(\sqrt{\bar{\alpha}_t}z_0+\sqrt{1-\bar{\alpha}_t}\epsilon,x,t)\rVert^2\Big]
\end{multline}
where $\epsilon\sim N(0,I_m)$ is the noise used to construct $z_t$, $\bar{\alpha}_t=\prod_{s=1}^t(1-\beta_s)$, and $\epsilon_\theta$ is a function trained to predict this noise given attributes $x$ and timestep $t$.

Our adaptation of the training algorithm is presented in Algorithm \ref{algo:train}. The noise model $\epsilon_\theta$ is a six-level convolutional U-Net with self-attention at the bottleneck. The noisy CAPE field $\sqrt{\bar{\alpha}_t}y + \sqrt{1 - \bar{\alpha}_t}\epsilon$ and attributes $x$ are concatenated along the channel axis before being passed to the noise model. The timestep $t$ is encoded via a sinusoidal embedding.

When evaluating our models, we found poor performance near the edges of the spatial domain, which we attribute to edge-padding applied to the field for convolutional filters. We leave a detailed study of the effects of padding for future research, and address this issue by training the DM on a larger spatial domain and cropping to the desired area after sampling.

Ensemble members are sampled via the standard reverse diffusion process. For each member, an initial random field is drawn from $N(0,I_m)$. The sample is iteratively denoised over $T$ iterations, using $\epsilon_\theta$ at each step to draw from the learned reverse transition $p_\theta(z_{t-1}\mid z_t)$, which includes a stochastic term whose variance decreases with $t$. This is repeated until the final forecast $z_0$ is generated. Our adaptation of the sampling algorithm from \citet{Ho2020} is presented in Algorithm \ref{algo:sample}.

\begin{algorithm}
\caption{Training algorithm from \citet{Ho2020} adapted to our setting. The noise model $\epsilon_\theta$ is a six-layer convolutional U-Net. The noisy CAPE field $\sqrt{\bar{\alpha}_t}y + \sqrt{1 - \bar{\alpha}_t}\epsilon$ and attributes $x$ are concatenated along the channel axis before being passed to the noise model, and the timestep $t$ is encoded via a sinusoidal embedding. }
\label{algo:train}
\begin{algorithmic}[1]
\REPEAT
    \STATE Sample CAPE field $y$ and corresponding attributes $x$ from dataset 
    \STATE Sample timestep $t \sim \text{Uniform}\{1, \ldots, T\}$
    \STATE Sample noise $\epsilon \sim N({0}, {I})$
    \STATE Take gradient descent step on
    \[
        \nabla_\theta \left\| \epsilon - \epsilon_\theta(\sqrt{\bar{\alpha}_t}{y} + \sqrt{1 - \bar{\alpha}_t}\epsilon,x, t) \right\|^2
    \]
\UNTIL{converged}
\end{algorithmic}
\end{algorithm}

\begin{algorithm}
\caption{Sampling algorithm from \citet{Ho2020} adapted to our setting. The pretrained U-Net $\epsilon_\theta$ is used to denoise the random sample.}
\label{algo:sample}
\begin{algorithmic}[1]
\STATE Sample noise ${z}_T \sim N({0}, {I})$
\FOR{$t = T, \ldots, 1$}
    \STATE ${\epsilon} \sim N({0}, {I_m})$ if $t > 1$, else ${\epsilon} = {0}$
    \STATE ${z}_{t-1} = \frac{1}{\sqrt{1-\beta_t}} \left( {z}_t - \frac{\beta_t}{\sqrt{1 - \bar{\alpha}_t}} \epsilon_\theta({z}_t, x, t) \right) + \sigma_t {\epsilon}$
\ENDFOR
\RETURN ${z}_0$
\end{algorithmic}
\end{algorithm}

\subsection{Two-Stage Training}
\label{sec:twostage}

\begin{sidewaysfigure*}[h]
 \noindent\includegraphics[width=45pc,angle=0]{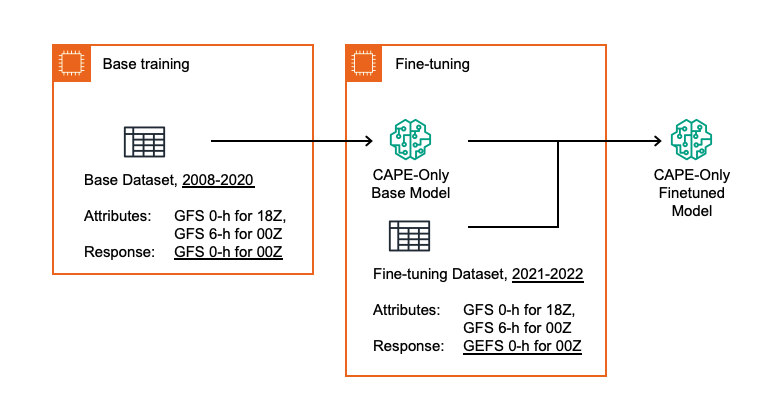}\\
 \caption{Diagram of how the two-stage pipeline is used to train the CAPE-Only Finetuned model. The base and finetuning datasets only differ in the response and time period. Base training takes 800 epochs and fine-tuning takes 200 epochs.} \label{fig:flow}
\end{sidewaysfigure*}

To fairly compare the AI models to GEFS we use the GEFS 0-h control forecast issued at 00Z as a proxy for the ground truth analysis. Ideally, this field would also serve as the response in our training set. However, GEFS v12 has only been operational for six years, providing insufficient historical data to properly train and evaluate a DM. To solve this, we first train our models on a base dataset that uses the GFS 0-h forecast as the response, then we fine-tune on a dataset that uses the GEFS 0-h control forecast as the response. The training of the CAPE-Only Finetuned model is depicted in Figure \ref{fig:flow}.

We construct a large base dataset of summertime forecasts from 2008 to 2020, using an extended summer defined as April 1 to September 30. Due to data availability issues, which are further discussed in Appendix A, we were unable to obtain data for approximately 5\% of days, resulting in a dataset with $n_{base}=2619$ days. This dataset has attributes from GFS and MERRA-2, with a response from GFS. We then construct the fine-tuning dataset, which also has GFS and MERRA-2 as attributes, but uses the GEFS 0-h forecast control member as the response. This smaller dataset contains summertime forecasts from 2021 and 2022, yielding $n_{ft}=366$ days. We use data from 2023 to 2025 to create a test set, which has the same attributes and response as the fine-tuning dataset. The base and fine-tuning datasets share the same input variables, only differing in the response and time period. 

The two stage training is performed by first initializing the model weights randomly and training for 800 epochs on the base dataset to produce a base model. We then produce a finetuned model by continuing training for 200 epochs on the fine-tuning dataset. No model weights are frozen during fine-tuning. We empirically validate our two-stage procedure by training three variants of the Aerosol GFS Base and Aerosol Finetuned models. The Aerosol GFS Base model does not use the two-stage pipeline and is only trained on the dataset with GFS response. The Aerosol GEFS Base model also does not use two-stage training, but is trained on the smaller dataset with GEFS as response. These two models will help evaluate whether the two-stage pipeline actually produces a better model. The Aerosol Partially Finetuned model uses the two-stage procedure, but with an abridged version of the fine-tuning dataset that only contains observations from 2021. This model helps to measure the importance of the size of the fine-tuning dataset. The three models are summarized in Table \ref{tab:models}. 

Training the CAPE-Only Base and Aerosol Base models for 800 epochs takes approximately 27 hours. Fine-tuning the models on the GEFS dataset for 200 epochs to produce the CAPE-Only Finetuned and Aerosol GFS Finetuned models takes an additional 40 minutes. The Aerosol GEFS Base model is only trained for 800 epochs and takes 3 hours. The Aerosol Partially Finetuned model takes 26 minutes for the 200 fine-tuning epochs. All models were trained on a single A40 NVIDIA GPU. Sampling an ensemble of $K=30$ members with $T=2000$ timesteps takes approximately four minutes for the CAPE-Only models and five minutes for the slightly larger aerosol models. When sampling ensembles for all the days in our test set, we use multiple GPUs to complete the task in parallel. 

\subsection{Classifier-free Guidance and Spread-Skill Ratio}\label{sec:cfg}

Training a DM is equivalent to learning a Stein score function $\nabla_{z_t} p_\theta(z_t \mid x)$. Sampling is then accomplished by using the score function to produce a forecast $z_0$ that has a high likelihood given attributes $x$. The score function can be (trivially) decomposed 
\begin{equation}
    \nabla_{z_t} p_\theta(z_t \mid x) = \gamma\nabla_{z_t} p_\theta(z_t \mid x)+(1-\gamma)\nabla_{z_t} p_\theta(z_t)
\end{equation}
as the linear combination of a conditional score $p_\theta(z_t \mid x)$ and an unconditional score $p_\theta(z_t)$. In classifier-free guidance, both score functions are learned implicitly and the tuning parameter $\gamma$ is used to control how closely the model relies on the attributes $x$ \citep{Ho2021}. When $\gamma \approx1$ the model relies heavily on the attributes and exhibits less variation in its forecasts. When $\gamma \approx 0$, the model gives less weight to the attributes and exhibits more variation in its forecasts.

Both the CAPE-Only and aerosol models are trained to support guidance. However, we replace the unconditional score $p_\theta(z_t)$ with a partial conditional score $p_\theta(z_t\mid x')$, where $x'$ is a subset of the features in $x$. In our model we omit the GFS 6-h forecast so that we control how closely the model follows the deterministic forecast. The partial conditional score is trained implicitly by masking the GFS 6-h forecast at random iterations.

We found that we could adjust $\gamma$ to control the spread-skill ratio (SSR) of the forecasts. The SSR is the ratio
\begin{equation}
    \text{SSR} = \frac{\text{Spread}}{\text{RMSE}}
\end{equation}
of the spread, or standard deviation, of the forecast ensemble to the RMSE. The spread and RMSE are averaged across the spatial domain and are formally defined in Appendix B. Forecasters often seek to calibrate models such that the $\text{SSR}=1$ \citep{Fortin2014}.

\begin{figure}[h]
 \noindent\includegraphics[width=19pc,angle=0]{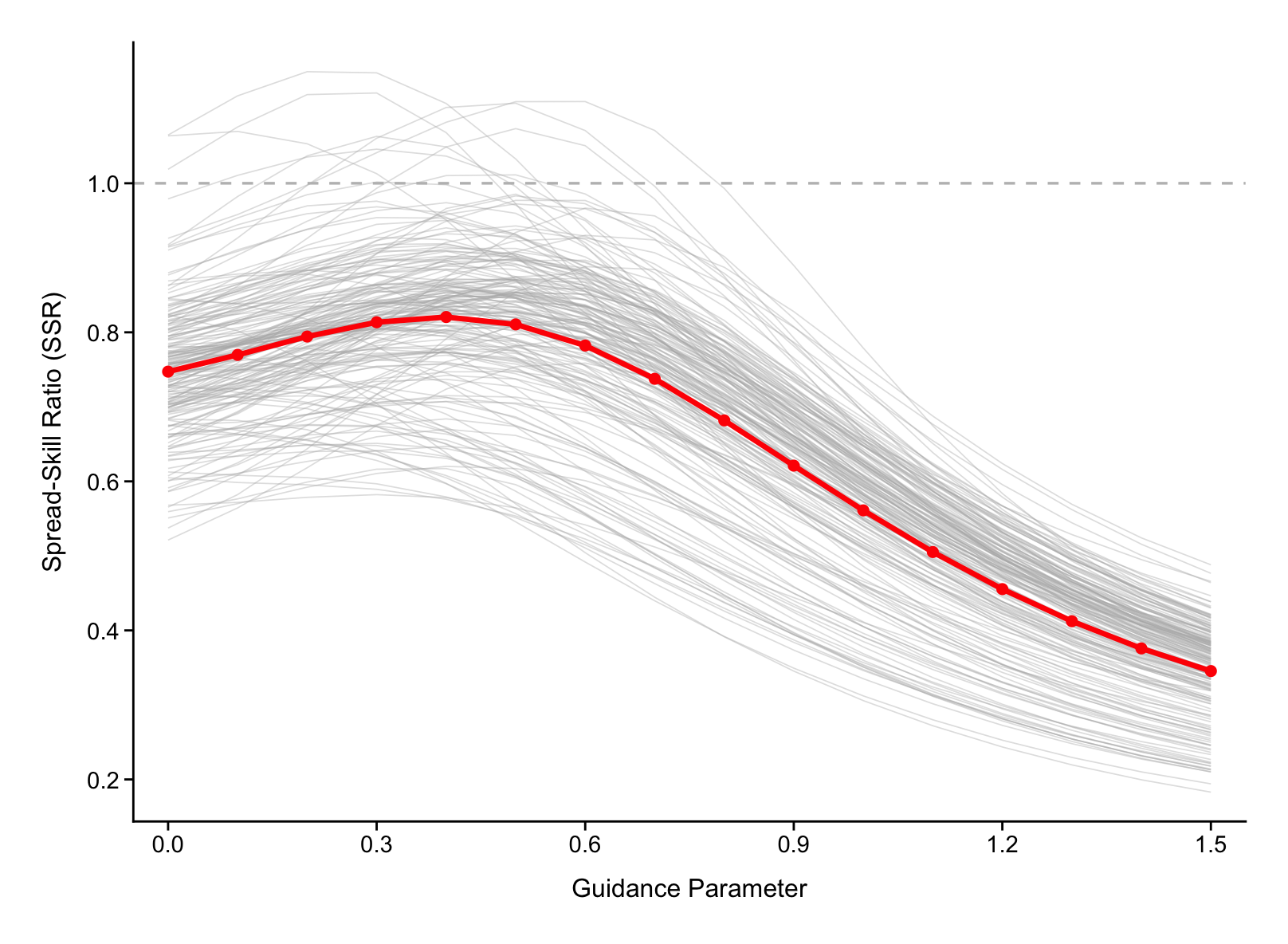}\\
 \caption{Spread-skill ratio (SSR) vs guidance for forecasts produced with the Aerosol Finetuned model. To minimize computational cost $T'=100$ timesteps are used instead of $T=2000$. Guidance parameters range from $\gamma=0$ to $\gamma=1.5$ in increments of $0.1$. Each gray line corresponds to the relationship for one day in the summer of 2022. The red line is the average across all days.} \label{fig:ssrg}
\end{figure}

\begin{figure}[h]
 \noindent\includegraphics[width=19pc,angle=0]{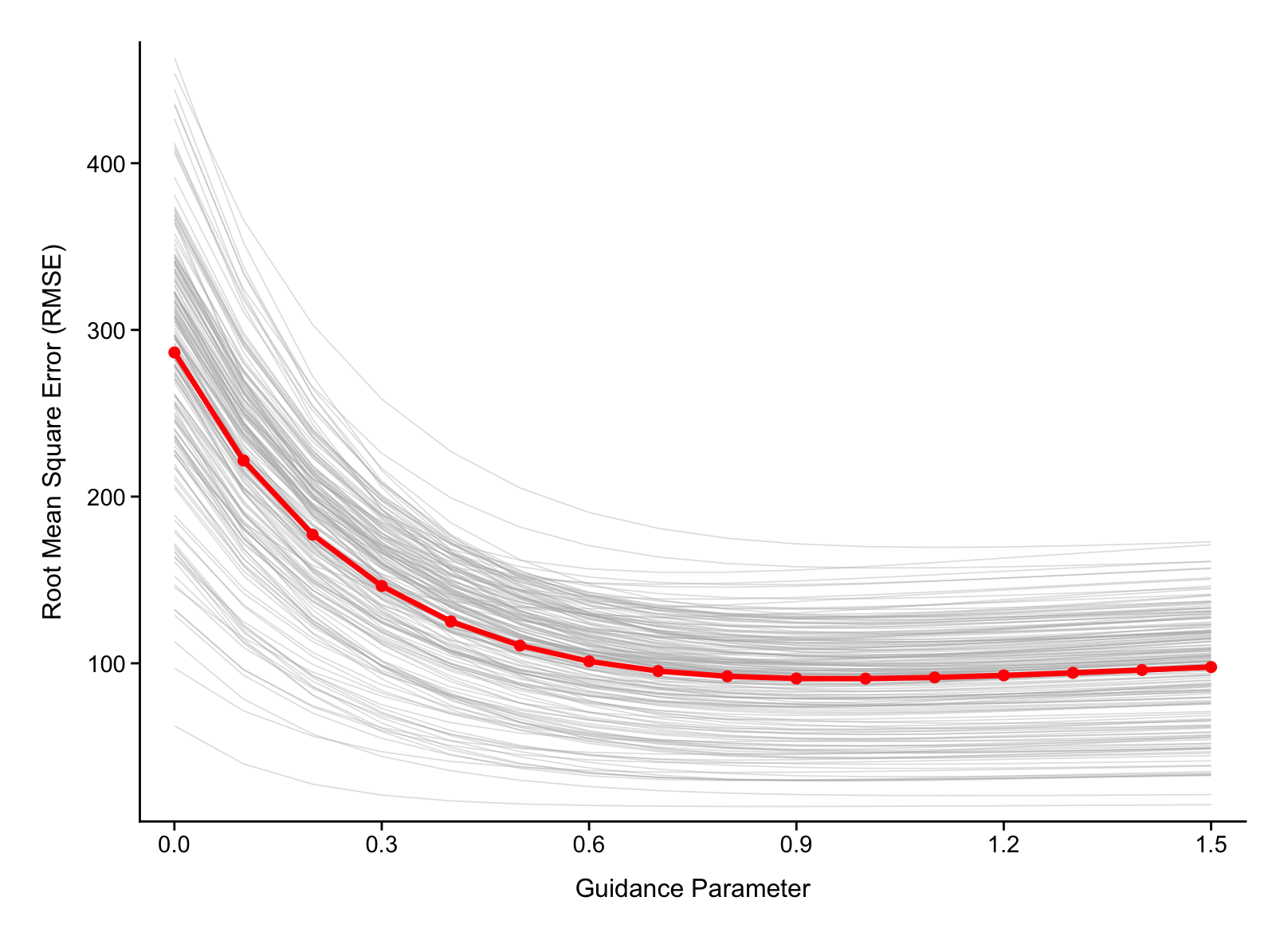}\\
 \caption{RMSE vs guidance for forecasts produced with the Aerosol Finetuned model. To minimize computational cost $T'=100$ timesteps are used instead of $T=2000$. Guidance parameters range from $\gamma=0$ to $\gamma=1.5$ in increments of $0.1$. Each gray line corresponds to the relationship for one day in the summer of 2022. The red line is the average across all days.}\label{fig:rmseg}
\end{figure}

\begin{figure}[h]
 \noindent\includegraphics[width=19pc,angle=0]{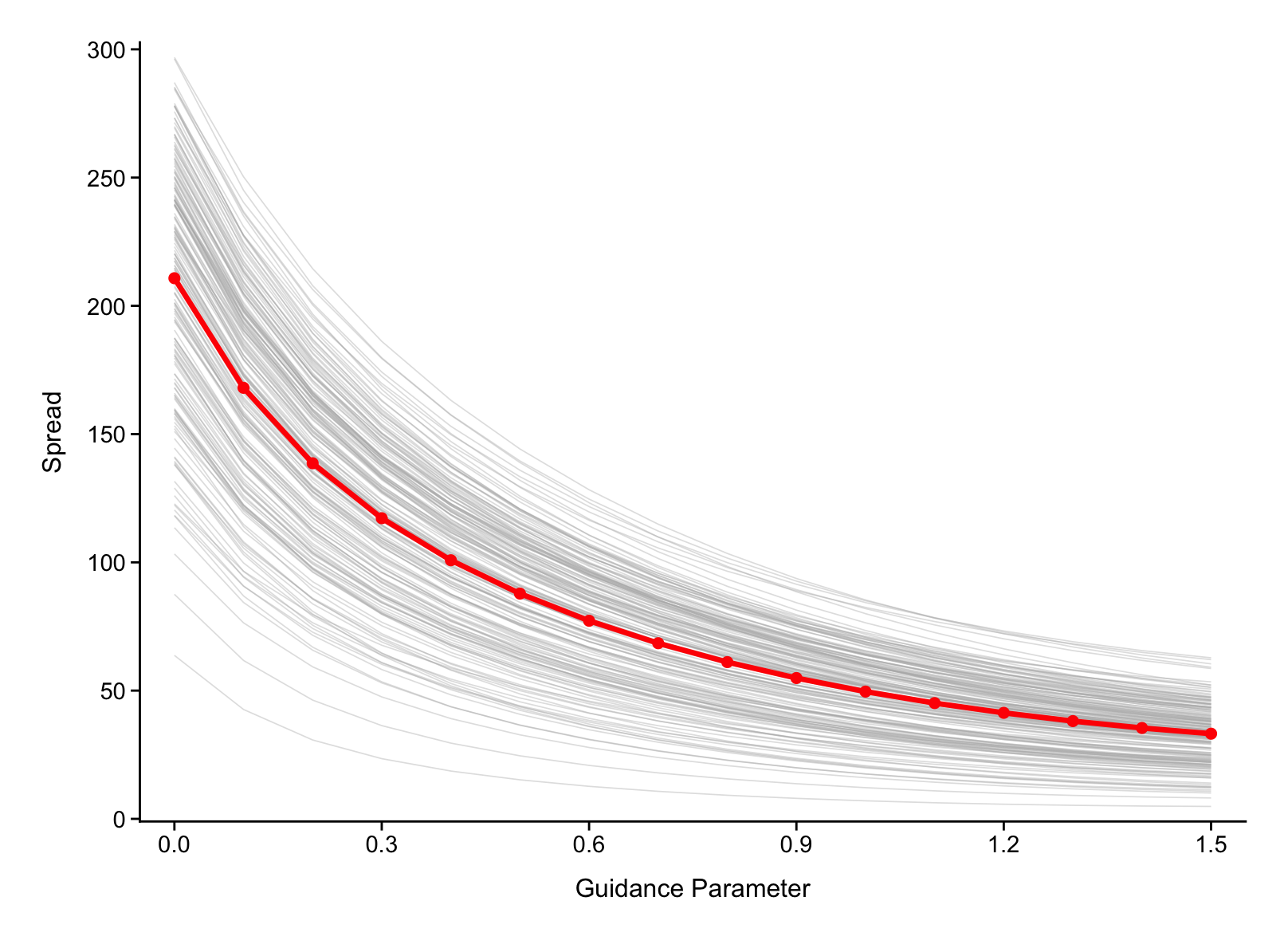}\\
 \caption{Spread vs guidance for forecasts produced with the Aerosol Finetuned model. To minimize computational cost $T'=100$ timesteps are used instead of $T=2000$. Guidance parameters range from $\gamma=0$ to $\gamma=1.5$ in increments of $0.1$. Each gray line corresponds to the relationship for one day in the summer of 2022. The red line is the average across all days.}\label{fig:spreadg}
\end{figure}

We empirically determined the relationship between the guidance parameter and SSR by producing forecasts with the Aerosol Finetuned model for each day in the summer of 2022 for sixteen different values of $\gamma$ and recording the SSR. Sampling ensembles over the summer season multiple times is computationally expensive, even when using multiple GPUs in parallel. We address this by using an approximation to the sampling algorithm with $T'=100$ timesteps instead of $T=2000$. The resulting model is still a DDPM, but uses a coarser time discretization. This results in fast sampling of approximately ten seconds per ensemble, but also slightly worse forecasts. A plot of SSR against guidance is presented in Figure \ref{fig:ssrg}, with the corresponding RMSE and Spread in Figure \ref{fig:rmseg} and Figure \ref{fig:spreadg}.

Forecasts with $\gamma < 0.5$ display low levels of skill, with the RMSE increasing the smaller the guidance parameter becomes. This behavior is expected since the model needs the GFS 6-h forecast to produce the ensemble, and at small $\gamma$ values it is being ignored. For $\gamma > 0.5$ the SSR decreases as the spread remains flat and the RMSE slightly increases. We determined that $\gamma = 0.6$ produces forecasts with high skill and a reasonable SSR.

\subsection{Metrics}

We evaluate the models using CRPS, RMSE, and Brier score. CRPS is a proper scoring rule
\begin{align}
    \text{CRPS}(F,y)&=\int_{-\infty}^\infty (F(s)-1\{y\leq s\})^2 ds\\
    &= \mathbb{E}_{Y\sim F}\lvert Y-y\rvert-\frac{1}{2}\mathbb{E}_{Y,Y'\sim F}\vert Y-Y'\rvert
\end{align}
that measures how well the predictive distribution $F$ matches the true outcome $y$, where $Y,Y'\sim F$ \citep{Gneiting2014}. The CRPS of the different models are compared using a skill score
\begin{equation}
    \text{CRPSS} = 1-\frac{\text{CRPS}_{\text{forecast}}}{\text{CRPS}_{\text{GEFS}}}
\end{equation}
relative to the GEFS forecasts. For a deterministic forecast, like GFS, the CRPS simplifies to mean absolute error. The RMSE measures how well the mean of the predictive ensemble matches the ground truth. Since our prediction is an ensemble, we use a bias-corrected RMSE and then calculate a RMSE skill-score in a similar fashion to $\text{CRPSS}$. 

Accurately predicting extreme CAPE is important for forecasting severe weather events. We use a bias-corrected Brier score \citep{Ferro2013} to compare the ability of forecasts to predict extreme CAPE values. The Brier score is a strictly proper scoring rule for evaluating the performance of categorical forecasts. The forecasts are converted to binary predictions of whether CAPE exceeds the $99\%$, $99.9\%$, or $99.99\%$ quantile of CAPE values across the training dataset, which corresponds to $2462$ J/kg, $3799$ J/kg, and $4846$ J/kg. We then calculate a Brier skill score for each forecast
\begin{equation}
    \text{BSS} = 1 - \frac{\text{Brier}_{\text{forecast}}}{\text{Brier}_{\text{clim}}}
\end{equation}
which compares the performance of the forecast to a climatology. To compare two models, we calculate the average difference in Brier skill score across all forecasts. 

The metrics are evaluated on a test set of summertime CAPE fields from 2023 to 2025, consisting of $n_{test}=549$ days. We produce ensembles with $K=30$ members, although all our models are capable of producing larger ensembles at minimal cost. We calculate the average skill score across days and provide $95\%$ confidence intervals using the stationary bootstrap \citep{Politis1994}. Additional details on metric calculation are provided in Appendix B.

\section{Results}
\label{sec:results}

The CRPS and RMSE skill scores of the CAPE-Only Finetuned, Aerosol Finetuned, and GFS models are presented in Table \ref{tab:crps_rmse_main}. The CAPE-Only Finetuned model significantly outperforms GEFS, with an average CRPS that is $20.4\%$ lower and an average RMSE that is $22.4\%$ lower than GEFS. The skill scores of the CAPE-Only model are also greater than GFS. All pairwise differences in Table \ref{tab:crps_rmse_main} are statistically significant, with confidence intervals for the mean differences presented in Appendix C. We expect GFS to underperform GEFS in CRPS, as it is a deterministic forecast that cannot capture model uncertainty.

The Aerosol Finetuned model significantly outperforms GEFS, with an average $24.9\%$ lower CRPS and $23.5\%$ lower RMSE. It also outperformed the CAPE-Only Finetuned and GFS models. These results suggest that incorporating aerosol information improves CAPE forecasts.

The Brier scores for the CAPE-Only Finetuned, Aerosol Finetuned, and GFS models are presented in Table \ref{tab:brier_main}. GFS does not perform significantly better than GEFS at predicting extreme CAPE values. The CAPE-Only Finetuned and Aerosol Finetuned models are significantly better than GEFS, but are not significantly better than one another. This suggests that DMs are skilled at enhancing a forecast's ability to predict extreme events, but that the additional aerosol information had little impact. It is possible that the model was unable to learn the relationship between aerosols and extreme CAPE.

The CRPS and RMSE skill scores of the models discussed in Section 3\ref{sec:twostage} are presented in Table \ref{tab:crps_rmse_twostage}. The Aerosol GEFS Base model performs only marginally better than GEFS, which is expected given the small training set. The Aerosol GFS Base model performs significantly better than GEFS. This performance is further improved by fine-tuning, as shown by the performance of the Aerosol Finetuned model. However, we find that fine-tuning requires a sufficiently large fine-tuning dataset to achieve improvement, as the Aerosol Partially Finetuned model, which uses a one-year fine-tuning dataset, does not outperform the Aerosol GFS Base model.

The Brier scores of the four models do not differ significantly, as seen in Table \ref{tab:brier_twostage}. This suggests that enhancing GEFS to better predict extreme values is relatively easy, while improving the overall skill and uncertainty quantification is more challenging. 

\begin{table*}[t]
\caption{CRPS and RMSE skill scores for CAPE-Only Finetuned, Aerosol Finetuned, and GFS forecasts with associated $95\%$ confidence intervals. Skill scores are computed relative to GEFS, so positive values indicate better performance than GEFS. For example, the Aerosol Finetuned model has an average CRPS 24.9\% lower than GEFS. Confidence intervals are calculated with the stationary bootstrap.}\label{tab:crps_rmse_main}
\begin{center}
\begin{tabular}{lccc}
\hline\hline
 & CAPE-Only Finetuned& Aerosol Finetuned & GFS\\
\hline
CRPSS & $20.4\pm2.0$ & $24.9\pm1.7$ & $-13.1\pm1.8$\\
RMSESS & $22.4\pm1.9$ & $23.5\pm1.7$ & $13.3\pm1.5$\\
\hline
\end{tabular}
\end{center}
\end{table*}

\begin{table*}[t]
\caption{The difference in Brier skill scores between each of the CAPE-Only Finetuned, Aerosol Finetuned, and GFS forecasts and GEFS for predicting CAPE values at the 99\%, 99.9\%, and 99.99\% quantile. The Brier skill score measures how much lower the forecast's Brier score is relative to a climatology. A positive difference in Brier skill scores indicates the model outperforms GEFS. Associated $95\%$ confidence intervals are calculated with the stationary bootstrap.}\label{tab:brier_main}
\begin{center}
\begin{tabular}{lccc}
\hline\hline
 & CAPE-Only Finetuned& Aerosol Finetuned & GFS\\
\hline
Brier SS 99\%&  $0.23\pm0.069$&$0.24\pm0.065$&$-0.0032\pm0.086$
\\
Brier SS 99.9\% &  $30.33\pm19.77$&$27.89\pm18.59$&$-1.52\pm12.36$
 \\
Brier SS 99.99\% &  $1390.541\pm916.49$&$1519.92\pm1007.93$&$143.02\pm850.41$
\\
\hline
\end{tabular}
\end{center}
\end{table*}

\begin{table*}[t]
\caption{ 
CRPS and RMSE skill scores for the Aerosol GEFS Base, Aerosol GFS Base, Aerosol Partially Finetuned, and Aerosol Finetuned forecasts with associated $95\%$ confidence intervals. Skill scores are computed relative to GEFS, so positive values indicate better performance than GEFS. For example, the Aerosol model has an average CRPS that  is $24.9\%$ lower than GEFS. Confidence intervals are calculated with the stationary bootstrap.
}\label{tab:crps_rmse_twostage}
\begin{center}
\begin{tabular}{lcccc}
\hline\hline
 & Aerosol GEFS Base & Aerosol GFS Base & Aerosol Partially Finetuned & Aerosol Finetuned\\
\hline
CRPSS & $4.1\pm3.1$ & $15.0\pm2.4$ & $14.1\pm2.3$ & $24.9\pm1.7$\\
RMSESS & $4.2\pm2.8$ & $15.9\pm2.3$ & $15.4\pm2.0$ & $23.5\pm1.7$\\
\hline
\end{tabular}
\end{center}
\end{table*}

\begin{table*}[t]
\caption{
The difference in Brier skill scores between each of the Aerosol GEFS Base, Aerosol GFS Base, Aerosol Partially Finetuned, and Aerosol Finetuned model forecasts and GEFS for predicting CAPE values at the 99\%, 99.9\%, and 99.99\% quantile. Brier skill score measures how much lower the forecast's Brier score is relative to a climatology. A positive difference in Brier skill scores indicates the model outperforms GEFS. Associated $95\%$ confidence intervals are calculated with the stationary bootstrap.
}\label{tab:brier_twostage}
\begin{center}
\begin{tabular}{lcccc}
\hline\hline
 & Aerosol GEFS Base & Aerosol GFS Base & Aerosol Partially Finetuned & Aerosol Finetuned\\
\hline
Brier SS 99\% &  $0.16\pm0.073$&$0.11\pm0.051$&$0.15\pm0.069$&$0.24\pm0.065$
\\
Brier SS 99.9\% &  $35.84\pm24.59$&$17.38\pm19.46$&$34.64\pm23.74$&$27.89\pm18.59$
\\
Brier SS 99.99\% &  $2112.25\pm1756.42$&$1541.51\pm1077.20$&$2100.20\pm1706.58$&$1519.92\pm1007.93$
\\
\hline
\end{tabular}
\end{center}
\end{table*}

\section{Model Interpretation}
\label{sec:interpetation}

The Aerosol Finetuned model outperforms the CAPE-Only Finetuned model, suggesting that it uses the additional information to produce more skillful forecasts. We use permutation feature importance (PFI) to assess which aerosols have the greatest impact on predicted CAPE. PFI measures the change in model performance when an input feature is removed \citep{Molnar2025}. Training separate models with each feature removed is computationally infeasible, so we instead implicitly remove feature information by permuting its values in the test set. This produces an attribute field that is no longer informative of CAPE. The PFI calculation is described in Algorithm \ref{alg:pfi}.

PFI is sensitive to correlations among input features. When multiple features are strongly correlated, their individual PFI scores may be attenuated. This occurs because the model can still reconstruct the permuted feature from its correlated counterparts, reducing the impact of permuting any single variable. In our dataset, black carbon and organic carbon AOD values are highly correlated, with a correlation coefficient of $\rho_{BC,OC} = 0.90$. We address this by grouping the variables and applying the same permutation to both simultaneously. We acknowledge this results in a loss of information, as black and organic carbon have different radiative properties. Sulfates are also correlated with black and organic carbon, with $\rho_{SU,BC}=0.36$ and $\rho_{SU,OC}=0.29$. However, we consider these correlations sufficiently small to treat sulfates as a separate variable.

We calculate the PFI for all input variables using CRPS as the performance metric. We repeat the PFI calculation five times with different permutations to construct confidence intervals. For computational reasons, we use $T'=100$ timesteps as we did in Section \ref{sec:cfg}.

The PFI for each input variable in the Aerosol Finetuned model is reported in Table \ref{tab:pfi}. The GFS 6-h forecast has the greatest impact on model performance, which is expected because the model is implicitly correcting the GFS forecast based on the other inputs. The second most important variable is the GFS model initialization. This suggests that the model uses current CAPE conditions to correct errors in the GFS forecast.

Among the AOD fields, black and organic carbon had the greatest impact on CAPE forecasts. Black carbon strongly absorbs solar radiation and has a positive direct radiative forcing, while organic carbon both absorbs and scatters radiation \citep{Johnson2019}. Black and organic carbon AOD can vary significantly across the spatial domain due to different sources, such as forest fires and anthropogenic emissions, causing their impacts to be more localized than those of other aerosols  \citep{Huang2012}. As a result, black and organic carbon AOD information can help the model adjust CAPE forecasts in specific regions.

Sulfate AOD also had an impact on CAPE forecasts. Primary sulfate aerosols have multiple sources, including fossil fuel combustion and forest fires, resulting in a non-uniform AOD spatial distribution across CONUS. Sulfates primarily scatter solar radiation \citep{Haywood2000}, and are known to have a significant impact on radiative forcing \citep{Kiehl1993}.

The sea salt AOD had a PFI that was an order of magnitude less than that of the black and organic carbon, and approximately half of the PFI of sulfates, but is statistically greater than zero. Similar to sulfates, sea salt aerosols almost entirely scatter light at the wavelengths associated with solar irradiance \citep{Tang1997}. Unlike sulfates, they are produced almost entirely by wind interaction with the ocean. Since our spatial domain consisted of CONUS, high sea salt AOD was rare in our dataset, which may have limited its impact on model predictions.

Dust was the only aerosol with a PFI that did not differ statistically from zero. Dust aerosols are less prevalent over North America than in other regions, such as Africa and Asia \citep{Ginoux2001}. Within CONUS, dust concentrations are highest along the Pacific coast, a region that is less likely to display thermodynamic instability. As a result, dust information may not help the AI model improve CAPE predictions.

\begin{algorithm}
\caption{Permutation Feature Importance algorithm from \citet{Molnar2025}. The CRPS calculation is described in Appendix B.}
\label{alg:pfi}
\begin{algorithmic}[1]
\STATE \textbf{Input:} Forecast model $\hat{f}$, test set attributes $\mathbf{X}=\{X_i\}_{i=1,\ldots,n_{test}}$ with $p$ attributes and $n_{test}$ days, test set response $\mathbf{Y}=\{Y_i\}_{i=1,\ldots,n_{test}}$.
\STATE \textbf{Output:} Feature importance scores $\{FI_j\}_{j=1}^{p}$
\STATE Sample ensembles $\{F_i\}_{i=1,\ldots,n_{test}}$ using $\hat{f}$ and $\mathbf{X}$.
\STATE Estimate the original model error:
\[
e_{\text{orig}} = \frac{1}{n_{\text{test}}} \sum_{i=1}^{n_{\text{test}}} \widehat{\text{CRPS}}(F_i,Y_i)
\]
\FOR{each feature $j \in \{1, \ldots, p\}$}
    \STATE Generate feature set $\mathbf{X}_{j}$ by permuting feature $j$ in $\mathbf{X}$
    \STATE Sample ensembles $\{F_i^{j}\}_{i=1,\ldots,n_{test}}$ using $\hat{f}$ and $X_j$
    \STATE Estimate permutation error:
    \[
e_{j} = \frac{1}{n_{\text{test}}} \sum_{i=1}^{n_{\text{test}}} \widehat{\text{CRPS}}(F_i^j,Y_i)
\]
    \STATE Compute feature importance:
    \[
     FI_j = e_{j} - e_{\text{orig}}
    \]
\ENDFOR
\RETURN $\{FI_j\}_{j=1}^{p}$
\end{algorithmic}
\end{algorithm}

\begin{table}[t]
\caption{The PFI of the six input fields to Aerosol Finetuned model. $95\%$ confidence intervals are calculated by repeating each PFI calculation five times with different permutations. Raw CRPS is the loss metric. Samples are generated using a coarse grid of $T'=100$ timesteps due to computational constraints.}\label{tab:pfi}
\begin{center}
\begin{tabular}{lc}
\hline\hline
 & PFI \\
\hline
GFS 6-h forecast & $215.7\pm0.8$ \\
GFS 0-h forecast & $67.6\pm1.55$ \\
Black + organic carbon & $0.186\pm0.05$ \\
Sulfates & $0.053\pm0.009$ \\
Sea salt & $0.028\pm0.006$ \\
Dust & $-0.003\pm0.003$ \\
\hline
\end{tabular}
\end{center}
\end{table}

\section{Conclusion}
\label{sec:conclusion}

We presented an AI framework that produces high-quality CAPE forecast ensembles from a single deterministic forecast. The framework is able to incorporate aerosol information as an input variable, even though its relationship with CAPE is complex. Since GEFS has limited historical availability, we introduce a two-stage training pipeline that leverages the much longer GFS forecast record to improve our AI model. We also develop a variant of classifier-free guidance to control the SSR of the resulting forecasts. Our models significantly outperform both the input GFS forecasts and GEFS. Finally, we calculated PFI scores for the input variables and determined that black carbon, organic carbon, and sulfates had the largest impact on the CAPE forecasts, while sea salt was less influential and dust was statistically insignificant.

Our forecast models have several limitations. Firstly, they are not yet capable of operational forecasting in their current implementation, and only produce forecasts for a specific time of day at a specific lead time. Secondly, while the success of the Aerosol models proves that adding aerosols can improve forecasts, it does not prove a causal link between aerosols and CAPE. Finally, if we assume a causal link exists, as numerous studies suggest, then the model does not explain the underlying mechanism.

There exist several possible extensions to this study, some of which would address the model limitations. First, we could expand the spatial domain and add more lead times to produce a model that is suitable for operational forecasting. A three-dimensional spatial domain could capture behavior across different atmospheric levels. Adding additional variables for input and output would also result in a more complete forecast. Finally, we could apply a causal framework to the forecasts to better understand the impact of aerosols. While this would have to be formally defined, the ability of a diffusion model to sample from the conditional predictive distribution is encouraging.

\clearpage

\datastatement

NOAA GFS (\url{https://registry.opendata.aws/noaa-gfs-bdp-pds/}) and GEFS (\url{https://registry.opendata.aws/noaa-gefs/}) historical forecasts are available through AWS. MERRA-2 data is available from NASA Earthdata (\url{https://disc.gsfc.nasa.gov/datasets/M2T1NXAER_5.12.4/summary}). All code is available online (\url{https://github.com/zjames12/cape-aerosols}).

\appendix[A] 
\label{appendix:dm}
\appendixtitle{Diffusion Model Details}

\subsection{Variance Schedules}
We define the forward process of our Markov chain as
\begin{equation}
    Z_{t}\mid Z_{t-1} \sim N(\sqrt{1-\beta_t}Z_{t-1}, \beta_tI_m)
\end{equation}
and the reverse process as
\begin{equation}
    Z_{t-1}\mid Z_t \sim N(\mu_\theta(Z_t,x,.t), \sigma_t^2I_m)
\end{equation}
The forward variance schedule $\{\beta_t\}_{t=1,\ldots,T}$ starts at $\beta_0=0.0001$ and increases linearly to $\beta_T=0.02$. The reverse variance schedule is defined as $\sigma^2_t=\frac{1-\bar{\alpha}_{t-1}}{1-\bar{\alpha}_t}\beta_t$, where $\bar{\alpha}_t=\prod_{s=1}^t(1-\beta_s)$.

\subsection{Training Configuration}

We trained our model with the AdamW optimizer \citep{Loshchilov2019} using batches of size 16. The learning rate followed a cosine schedule with 1000 warm-up steps.

\subsection{Missing Data}

We obtain historical GFS CAPE forecasts by downloading pgrb2.0p50 GRIB files. These files contain forecasts of less commonly used atmospheric variables at the $0.5^\circ$-resolution, and are available from the NOAA Open Data Dissemination through several sources, including AWS and Google Cloud. However, we were unable to download the GRIB files for 156 days between August 2008 and July 2018. This data issue appears to be unrelated to the weather conditions or forecasts on these days, and we decided to omit them from our training set.

\appendix[B]
\appendixtitle{Metrics}
\label{appendix:metrics}

We trained a forecast model that produces ensembles $\mathbf{F}^k=\{F_i^k\}_{i=1,\ldots,M}$, where $M=30$, for each location $k$ in spatial domain $S$. These forecasts try to predict the observed CAPE $Y^k$. Similar to \citet{Price2025}, we define weights $a_k$ as proportional to the area of the corresponding latitude-longitude grid cell, which becomes larger the closer the cell is to the equator. The weights are normalized across $S$ to have unit mean.

\subsection{Continuous Ranked Probability Score}

The continuous ranked probability score (CRPS) is a strictly proper scoring rule. The CRPS is a function of the cumulative distribution function (CDF) of the predictive distribution. We treat the ensemble members as a sample from the predictive distribution and use the unbiased estimator 

\begin{equation}
    \widehat{\text{CRPS}}(\mathbf{F}^k,Y^k) = \frac{\sum_{i=1}^M\lvert F_i^k-Y^k\rvert}{M}+\frac{\sum_{i,j}^M\lvert F_i^k-F_j^k\rvert}{2M(M-1)}
\end{equation}

For a deterministic forecast, such as GFS, we calculate the CRPS as
\begin{equation}
    \widehat{\text{CRPS}}(F^k,Y^k) = \lvert F^k-Y^k\rvert
\end{equation}
the absolute error.
The CRPS for a forecast is then the weighted average
\begin{equation}
\widehat{\text{CRPS}}_{\text{forecast}} = \frac{1}{\vert S \rvert}\sum_{k\in S}a_k\widehat{\text{CRPS}}(\mathbf{F}^k,Y^k)
\end{equation}
across the spatial domain.
\subsection{RMSE}

The mean square error (MSE) measures how well the mean of the ensemble $\overline{F}^k=\frac{1}{M}\sum_{i=1}^MF^k_i$ captures the process. We use a bias-corrected MSE
\begin{equation}
    \text{MSE}(\mathbf{F}^k,Y^k) = (\overline{F}^{k}-{Y^k})^2 - \underbrace{\frac{s^2_{k}}{M}}_{\text{bias correction}}
\end{equation}
where $s^2_k$ is the ensemble standard deviation. The correction is needed since we are using the mean of an ensemble, and if we assume that each $F_i^k$ is drawn from a distribution with finite mean $\mu$ then the naive estimator
\begin{align}
        \mathbb{E}[\text{MSE}(\overline{F}^{k},Y^k)] &= \mathbb{E}[(Y^k-\overline{F}^{k})^2]\\
        &= \mathbb{E}[(\overline{F}^{k})^2]-2\mu Y^k+({Y^k})^2\\
        &= Var[\overline{F}^{k}] + \mu^2-2\mu Y^k+({Y^k})^2\\
        &= Var[\overline{F}^{k}] + \text{MSE}(\mu,Y^k)\\
        &= \frac{Var[F_i]}{M} + \text{MSE}(\mu,Y^k)
    \end{align}
is biased. When calculating the MSE of a deterministic forecast we assume $Var[F_i]=0$ and we get back the MSE without bias correction.

The RMSE for the full forecast is calculated as the 
\begin{equation}
    \text{RMSE}_{\text{forecast}} = \sqrt{\frac{1}{\vert S \rvert}\sum_{k\in S}a_k{\text{MSE}}(\mathbf{F}^k,Y^k)}
\end{equation}
square root of the weighted average of MSE.

\subsection{Spread-Skill Ratio}

The spread is the standard deviation of the ensemble, which we define
\begin{equation}
    \text{Spread}_{\text{forecast}} = \sqrt{\frac{1}{\vert S \rvert}\sum_{k\in S}a_ks^2_k}
\end{equation}
as the square root of the weighted average variance. The spread-skill ratio is then naturally defined as
\begin{equation}
    \text{SSR} = \frac{\text{Spread}_{\text{forecast}}}{\text{RMSE}_{\text{forecast}}}
\end{equation}

\subsection{Brier score}

The Brier score is a strictly proper scoring rule for categorical predictions. To predict whether CAPE exceeds a value $q$, we use the binary forecast $F_i^{k,q} = \mathbbm{1}\{F_i^k>q\}$. This is then compared to the ground truth $Y^{k,q} = \mathbbm{1}\{Y^k>q\}$. We then calculate the bias-corrected Brier score
\begin{equation}
    \text{Brier}(\mathbf{F}^k,Y^k) = \left(\frac{\overline{F}_i^{k,q}}{M}-Y^{k,q}\right)-\frac{\overline{F}_i^{k,q}(M-\overline{F}_i^{k,q})}{M^2(M-1)}
\end{equation}
from \citet{Ferro2013}, where $\overline{F}_i^{k,q}=\frac{1}{M}\sum_{i=1}^MF_i^{k,q}$. We do not use the bias correction for deterministic forecasts. The Brier score for the forecast is then a weighted average
\begin{equation}
    \text{Brier}_{\text{forecast}} = \frac{1}{\vert S \rvert}\sum_{k\in S}a_k\text{Brier}(\mathbf{F}^k,Y^k)
\end{equation}

across the spatial domain $S$.

\appendix[C]
\appendixtitle{Additional Results}
\label{appendix:addres}

\begin{table*}[h]
\caption{Pairwise differences of CRPS and RMSE skill scores for the CAPE-Only Finetuned, Aerosol Finetuned, and GFS forecasts with associated $95\%$ confidence intervals. Skill scores are computed relative to GEFS. Confidence intervals are calculated with the stationary bootstrap.}\label{tab:crps_diffs}
\begin{center}
\begin{tabular}{lcc}
\hline\hline
 & CRPS Skill Score Difference & RMSE Skill Score Difference \\
\hline
Aerosol Finetuned - CAPE-Only Finetuned & $4.5\pm0.6$ &  $1.1\pm0.5$\\
CAPE-Only Finetuned - GFS & $33.5\pm1.5$  & $9.1\pm1.4$ \\
Aerosol Finetuned - GFS & $38.0\pm1.2$& $10.2\pm1.3$\\
\hline
\end{tabular}
\end{center}
\end{table*}

\bibliographystyle{ametsocV6}
\bibliography{references}

\end{document}